\documentclass[twocolumn,showpacs,,usortedaddress, superscriptaddress, preprintnumbers,amsmath,amssymb]{revtex4}

\usepackage{graphicx}
\usepackage{dcolumn}
\usepackage{gensymb}
\usepackage{bm}
\usepackage{graphicx}
\usepackage{gensymb}
\usepackage{amsfonts}
\usepackage{hyperref}
\usepackage{url}

\begin{document}
\title{On the heterogeneous character of the heartbeat instability in complex (dusty) plasmas}

\author{M.Y. Pustylnik}
\email{pustylnik@mpe.mpg.de}
\affiliation{Max-Planck-Institut f\"{u}r Extraterrestrische Physik, Giessenbachstrasse, 85741 Garching, Germany}
\author{A.V. Ivlev}
\affiliation{Max-Planck-Institut f\"{u}r Extraterrestrische Physik, Giessenbachstrasse, 85741 Garching, Germany}
\author{N. Sadeghi}
\affiliation{LIPhy, Universit\'{e} de Grenoble 1/CNRS, UMR 5588, Grenoble 38401, France}
\author{R. Heidemann}
\affiliation{Max-Planck-Institut f\"{u}r Extraterrestrische Physik, Giessenbachstrasse, 85741 Garching, Germany}
\author{S. Mitic}
\affiliation{Max-Planck-Institut f\"{u}r Extraterrestrische Physik, Giessenbachstrasse, 85741 Garching, Germany}
\author{H.M. Thomas}
\affiliation{Max-Planck-Institut f\"{u}r Extraterrestrische Physik, Giessenbachstrasse, 85741 Garching, Germany}
\author{G.E. Morfill}
\affiliation{Max-Planck-Institut f\"{u}r Extraterrestrische Physik, Giessenbachstrasse, 85741 Garching, Germany}

\date{}

\begin{abstract}
A hypothesis on the physical mechanism generating the heartbeat instability in complex (dusty) plasmas is presented. It is suggested that the instability occurs due to the periodically repeated critical transformation on the boundary between the microparticle-free area (void) and the complex plasma. The critical transformation is supposed to be analogous to the formation of the sheath in the vicinity of an electrode. The origin of the transformation is the loss of the electrons and ions on microparticles surrounding the void. We have shown that this hypothesis is consistent with the experimentally measured stability parameter range, with the evolution of the plasma glow intensity and microparticle dynamics during the instability, as well as with the observed excitation of the heartbeat instability by an intensity-modulated laser beam, tuned to the atomic transition of a working gas.
\end{abstract}

 \maketitle

\section{Introduction}
Instabilities in electronegative discharges are known since a long time \cite{Emeleus}. The so-called attachment instabilities in glow dc discharges \cite{UFN}, as well as in capacitively-coupled \cite{Hollenstein} and inductively-coupled \cite{Tuszewski} rf discharges were observed. In the latter case the instability was found to be accompanied by the E-H transition [\onlinecite{Lieberman},\onlinecite{Chabert}]. All these instabilities reveal themselves as relatively low-frequency (1-100 kHz) self-excited oscillations of electrical and optical parameters of the discharge. Variation of optical emission has a global character, i.e., the emission changes in phase in the whole discharge. Therefore a global description implying homogeneous distribution of all plasma species is sufficient for understanding of such instabilities.

Complex (dusty) plasmas can also be considered as plasmas with very heavy negative ions. However, due to the macroscopic nature of microparticles, complex plasmas exhibit two features which make them remarkably distinct from regular electronegative plasmas: (i)~The charge of a microparticle is determined by the dynamic balance of the ion and electron fluxes on its surface \cite{Smy}, which represents therefore a volumetric sink for a plasma. (ii)~The ion drag force, acting on microparticles, becomes important, leading to the formation of so-called voids, i.e. microparticle-free regions [\onlinecite{Mugravity}-\onlinecite{Goedheer3}]. Such plasmas can no longer be treated as homogeneous.

In the first microgravity experiments with complex plasmas, a spontaneous periodic contraction of the void boundary was reported \cite{HBDisc}. Because of its characteristic appearance as well as due to very low repetition frequency (from single contractions to hundreds Hz), this phenomenon was termed "heartbeat" instability [\onlinecite{BoufendiHB1}-\onlinecite{pk3+}]. Several years after the discovery, it was shown in ground-based experiments that the heartbeat instability cannot be reduced to the dynamics of microparticles only \cite{BoufendiHB1, BoufendiHB}. The global characteristics of the discharge (e.g., rf current) were found to be modulated with the low frequency of the instability. High-speed real time imaging of the discharge showed that the contraction of the void is preceded by a steep increase of the glow inside it. Physical mechanisms leading to such a behavior of the plasma are not understood until now. Recent review of extensive data from the microgravity experiments, conducted on board of the International Space Station \cite{HeidemannHB}, also did not bring any new insights into the understanding of this phenomenon.

\begin{figure}[tb]
\centering
  \includegraphics [width=3.2in]{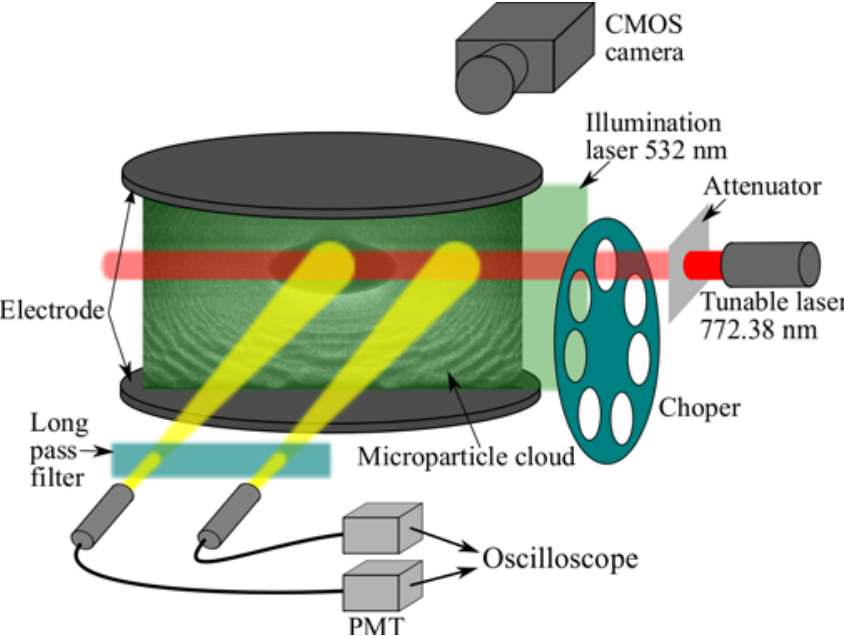}
  \caption{Experimental setup for the investigations of the heartbeat instability. Microparticles are levitated in the discharge volume with the help of the thermophoretic force. A $532$~nm laser is used for the illumination of the microparticles. A $772$~nm tunable laser, modulated by a chopper, provides additional means of exciting the heartbeat instability. Attenuators of different transparency are used to vary the tunable laser beam power. Two photomultiplyers are recording time-resolved emission intensity of the plasma. Longpass filter in front of them rejects the scattered light of $532$~nm illumination laser. A CMOS camera records the dynamics of microparticles.}
  \label{Fi:STP}
\end{figure}

 In this paper we suggest a hypothesis on the physical mechanism of the heartbeat instability in complex (dusty) plasmas. This hypothesis employs both distinct features of complex plasmas \textendash ~heterogeneity and the presence of a volumetric sink of plasma \textendash ~and does not contradict to any available experiments. Under this approach the driving mechanism of the heartbeat instability is the formation of a sheath at the void boundary, leading to the abrupt change in the profile of the electrostatic potential.

\section{Experimental setup}
Our experiments were performed in the PK-3+ chamber \cite{pk3+}, which is a symmetrically driven parallel plate capacitively coupled rf reactor (see Fig.~\ref{Fi:STP}). The discharge was created by applying $13.56$~MHz voltage to the disc-shaped electrodes of $6$~cm in diameter, separated by a $3$~cm gap. As a working gas we used argon in a pressure range of $p$ = $10-50$~Pa.

Melamine formaldehyde plastic spheres with the diameter $2a$ = $1.95$~$\mu$m were injected into the discharge and levitated in the vicinity of the bottom electrode. By heating up the bottom electrode and thus compensating the gravitational force by thermophoresis \cite{Rothermel} a significant part of the discharge volume was filled with microparticles, forming a void almost in the center of the chamber. The temperature difference between the electrodes was $19$~K. The microparticles were illuminated by a vertical sheet of green (532 nm) laser light and observed from the side with a CMOS video camera at a frame rate of $1000$~fps.

 At two radial positions \textendash ~one inside the void and the other outside it, on the periphery of the microparticle cloud (see Fig.~\ref{Fi:STP}) \textendash ~ the light from the plasma was collected by two lenses and by means of optic fibers transmitted to the two respective photomultiplier (PMT) modules with an amplifier bandwidth of $200$~kHz.

Along with the observation of the sporadically triggered heartbeat instability, we used a tunable diode laser to excite it "on purpose". The wavelength of the laser was set to the center of the Doppler profile of $\lambda$ = $772.38$~nm spectral line of argon. This light is resonantly absorbed by argon atoms in the 1s$_5$ metastable state, causing the excitation into the 2p$_7$ radiative state. The $2.4$~mm diameter cylindrical beam from the tunable laser with the maximal power of $8$~mW was modulated by a mechanical chopper and then horizontally driven through the chamber. It was possible to move the beam vertically. The laser power entering the chamber was varied by attenuating the beam with the neutral density filters. The same laser can be used for measuring the number density of 1s$_5$ metastable argon atoms by optical absorption in a manner similar to that of Ref. \onlinecite{TDLAS}.

\section{Results}
\subsection{Reproducibility}
A special note should be made regarding the reproducibility of the results. We were unable to get rid of the uncontrolled drift of the instability parameters. Significant change in, $e.g.$, frequency of the heartbeat, sometimes observed within $10-15$~min interval, while both discharge power and pressure were kept constant. Therefore, in each series of measurements we tried to keep the measurement time as short as possible, and always double checked that the observed effect is mainly due to the variation of the control parameter and not due to the degradation of the discharge conditions. \begin{figure}
\centering
  \includegraphics [width=3.in]{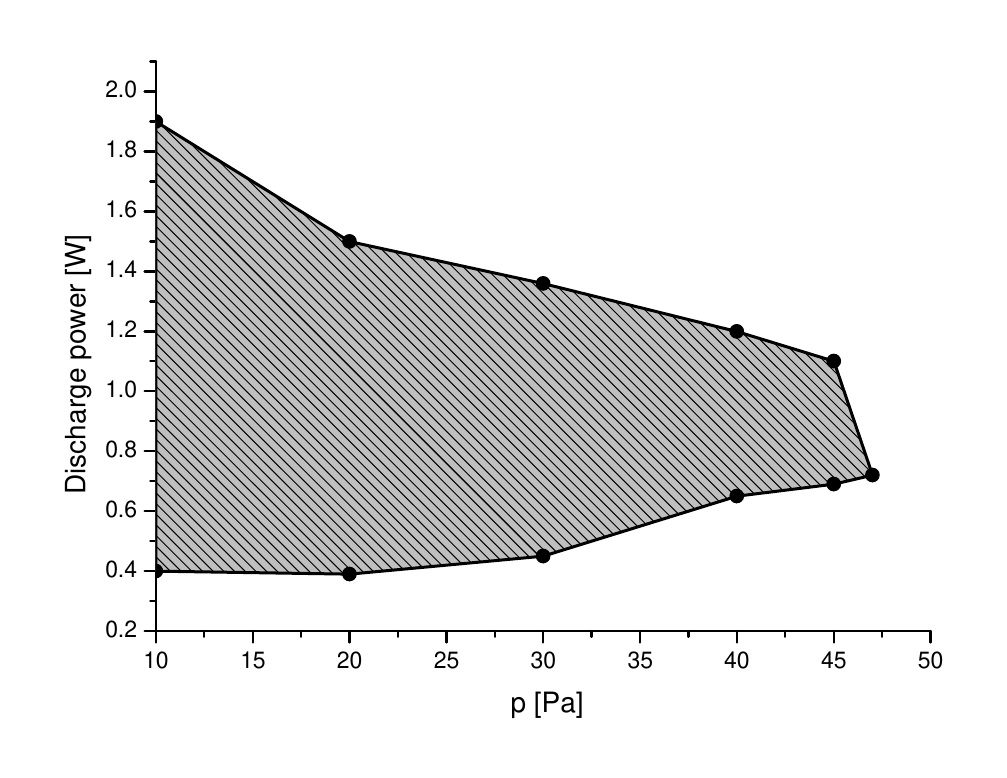}
  \caption{Example of the stability diagram, showing the parameter range where the heartbeat instability is triggered (inside the dashed area) for a given number of levitating microparticles. Circles represent the measured stability boundary which can vary significantly between different experimental series.}
  \label{Fi:HBPD}
\end{figure}
\subsection{Self-excited heartbeat instability}
The presence of the instability was found to be determined by a combination of three parameters: pressure, discharge power and number of microparticles. An example of a 2D stability map (at fixed number of microparticles) is shown in Fig.~\ref{Fi:HBPD}. Obviously, there exists a threshold pressure (47 Pa in the shown example), above which no instability can be excited at any discharge power. At lower pressures the instability can be triggered in a certain range of discharge powers, which broadens as the pressure is reduced. Below 10 Pa the discharge with microparticles cannot be sustained. For given pressure and power the instability sets in easier when the number of microparticles is larger.

\begin{figure}
\centering
  \includegraphics [width=3.in]{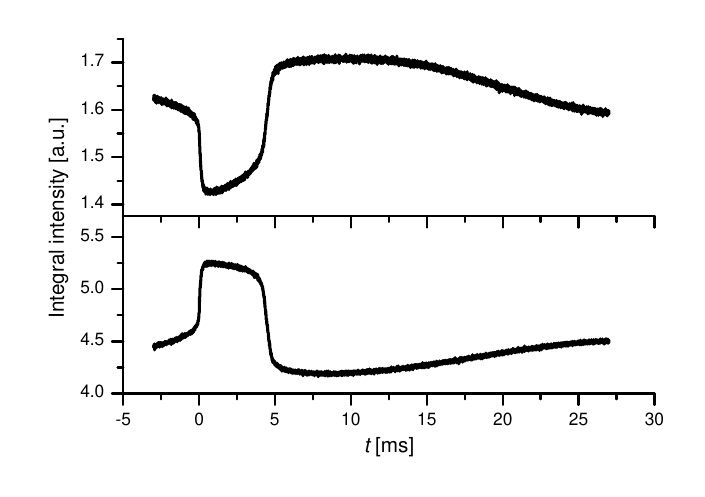}
  \caption{Evolution of the integral light intensity measured during the self-excited heartbeat instability (a) outside the void and (b) inside the void. The curves are obtained by averaging over 200 consecutive cycles of the instability.}
  \label{Fi:SE}
\end{figure}

In Fig.~\ref{Fi:SE} the evolution of the intensity of the plasma emission inside and outside the void is illustrated. To reduce the noise curves are averaged over 200 consecutive heartbeat events. We also made sure that two instability cycles never occurred within the time span of Fig.\ref{Fi:SE}.
Figure~\ref{Fi:SE} confirms the earlier observations \cite{BoufendiHB, HeidemannHB}. During the void-collapse phase of the heartbeat instability the glow intensity first rapidly (about $0.2$~ms in our case) increases inside the void and simultaneously decreases outside. Then both intensities exhibit a plateau for several ms. This is followed by an overshoot phase, when the intensities go above and below the average level outside and inside the void, respectively. Relaxation of this overshoot lasts until the next cycle of the instability starts. We would like to stress here that the intensities inside and outside the void oscillate remarkably out of phase.
\subsection{Laser excitation of the heartbeat instability}
Before performing the laser excitation of the heartbeat instability the system was brought to a stable condition. This was obtained by decreasing the discharge power 0.1-0.2 W below the lower onset threshold (which was observed between $0.8$ and $1.0$~W for the laser excitation experiments). Then, by shining an intensity-modulated light from a laser tuned to 2p$_7\to$ 1s$_5$ transition of argon we were able to observe the heartbeat instability \cite{Video}. We note, that the laser excitation of the heartbeat instability was performed in a dedicated series of experiments, conducted at $p=20$~Pa. This explains a significant deviation of the lower power threshold of the instability from that in Fig.~\ref{Fi:HBPD}.

\begin{figure}
\centering
  \includegraphics [width=3.in]{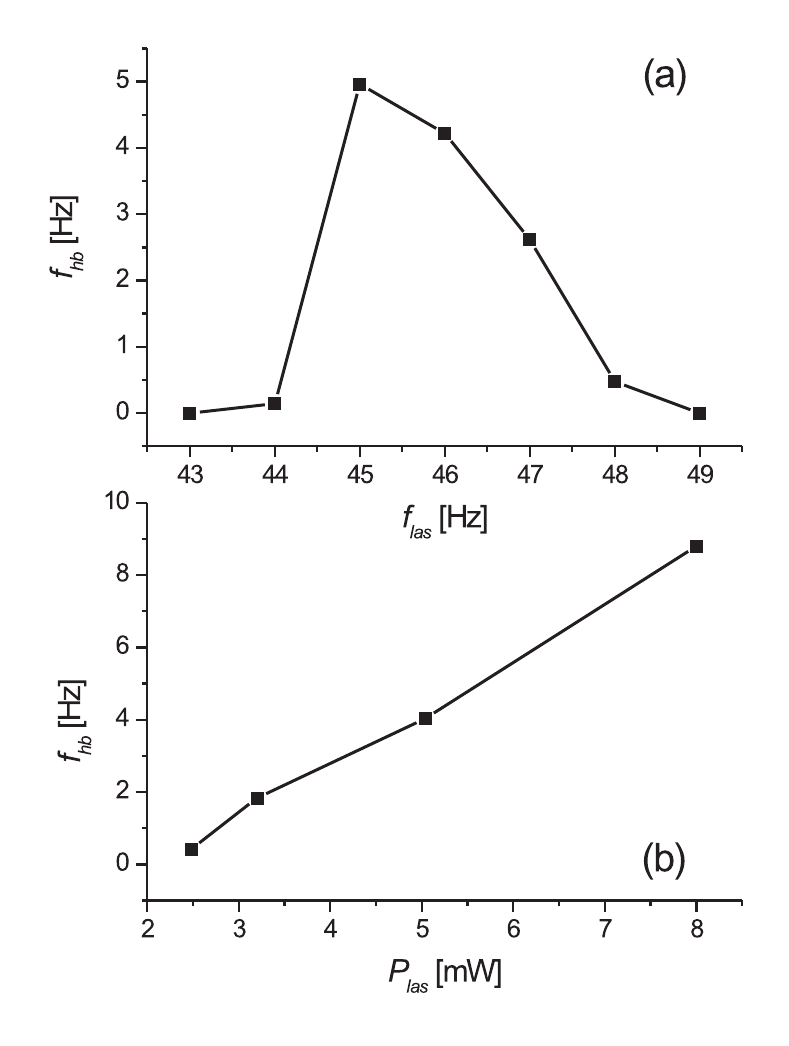}
  \caption{Dependence of the frequency of the laser-excited heartbeat instability, $f_{hb}$, on (a) the frequency of the laser modulation $f_{las}$ and (b) the tunable laser power $P_{las}$. The results are for (a) $P_{las}=3.5$~mW and (b)$f_{las}=55$~Hz. Laser beam crosses the center of the void.}
  \label{Fi:LEHBDep}
\end{figure}

The excitation of the heartbeat instability by a modulated beam of a tunable laser exhibits resonance properties. It was always observed in quite a narrow band of the beam modulation frequency $f_{las}$ around $40-60$~Hz. Example of a resonance curve is shown in Fig.~\ref{Fi:LEHBDep}a. We see that the frequency of the heartbeat, $f_{hb}$, can be used as a measure of "strength" of the instability. The laser-excited heartbeat instability becomes stronger with the increase of laser power, $P_{las}$ (Fig.~\ref{Fi:LEHBDep}b).

\begin{figure}
\centering
  \includegraphics [width=3.in]{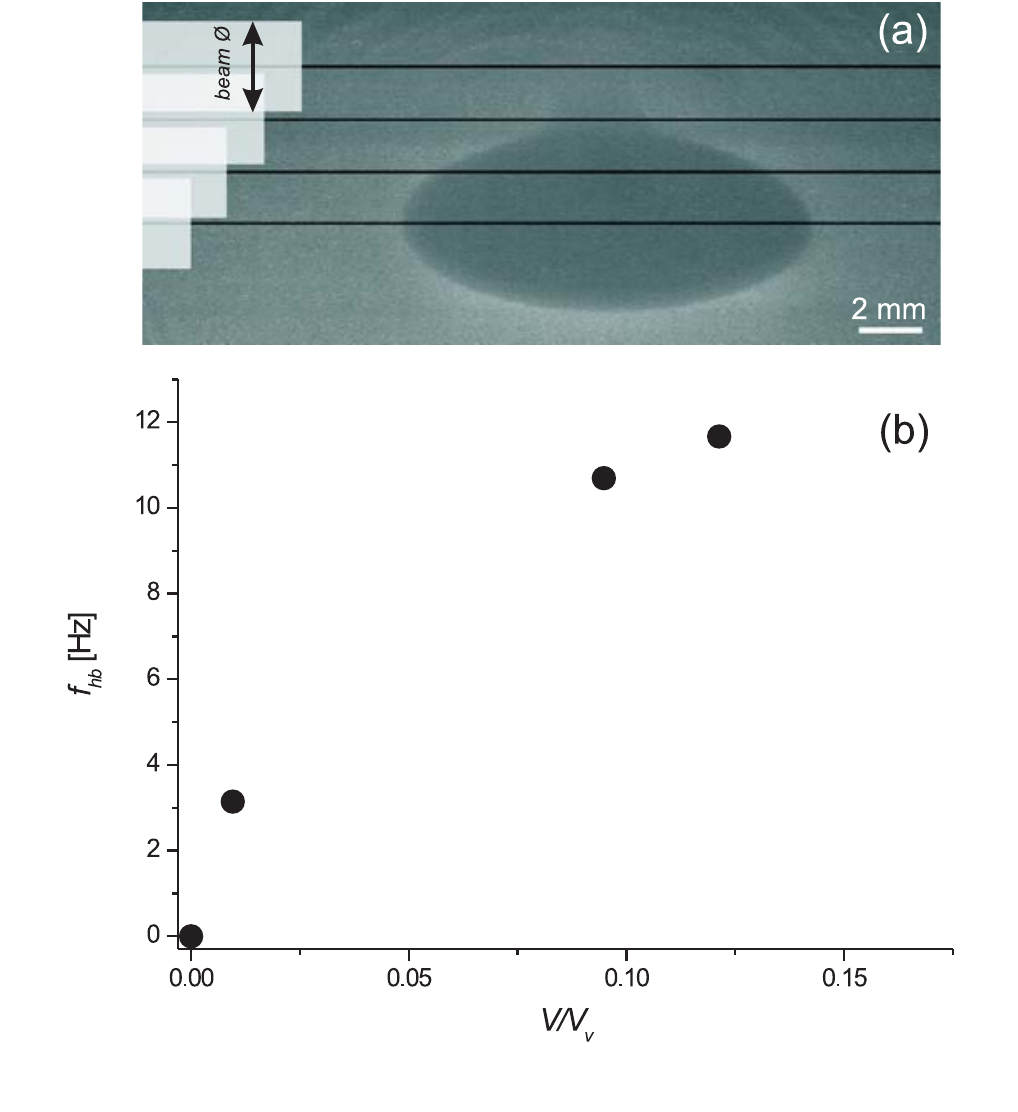}
  \caption{(a) Positions of the laser beam with respect to the void, at which (b) the dependence of the frequency of the laser-excited heartbeat instability, $f_{hb}$, on the fraction of the void volume occupied by the laser beam, $V/V_v$, was investigated. The void of volume $V_v$ is approximated as an oblate spheroid with the large and small semiaxes $6.5$ and $2.6$ mm respectively. Laser beam is a cylinder with the diameter of $2.4$ mm, $V$ is  defined as a volume of intersection of the laser beam and the void. For the shown example, $P_{las}$ = $8$ mW and $f_{las}$ = $50$~Hz. Movies of the microparticle cloud with the laser beam in the center of the void ($V/V_v=0.12$) and for the laser beam outside the void ($V/V_v=0$) can be viewed in Ref. [\onlinecite{Video}].}
  \label{Fi:HBh}
\end{figure}

The heartbeat instability strength depends also on the position of the laser beam \cite{Video}. Figure~\ref{Fi:HBh} shows the dependence of the frequency of the heartbeat instability on the fraction of the void volume occupied by the laser beam. The heartbeat frequency is maximal if the laser hits the center of the void, and decreases as the laser moves out towards the periphery of the cloud.

The heartbeat instability was also found to disappear if the laser is detuned from the argon atomic transition.

\section{Discussion}
\subsection{Critical transition at the void boundary}
The importance of the boundary between the void and surrounding complex plasma is substantiated by three experimental facts: (i) the instability never occurs without a void, (ii) variations of the intensity of the plasma glow are opposite in phase inside and outside the void, (iii) laser excitation of the instability is only possible if the plasma parameters are modulated \textit{inside} the void.
\begin{figure}[t!]
\centering
  \centerline{\includegraphics [width=3.in]{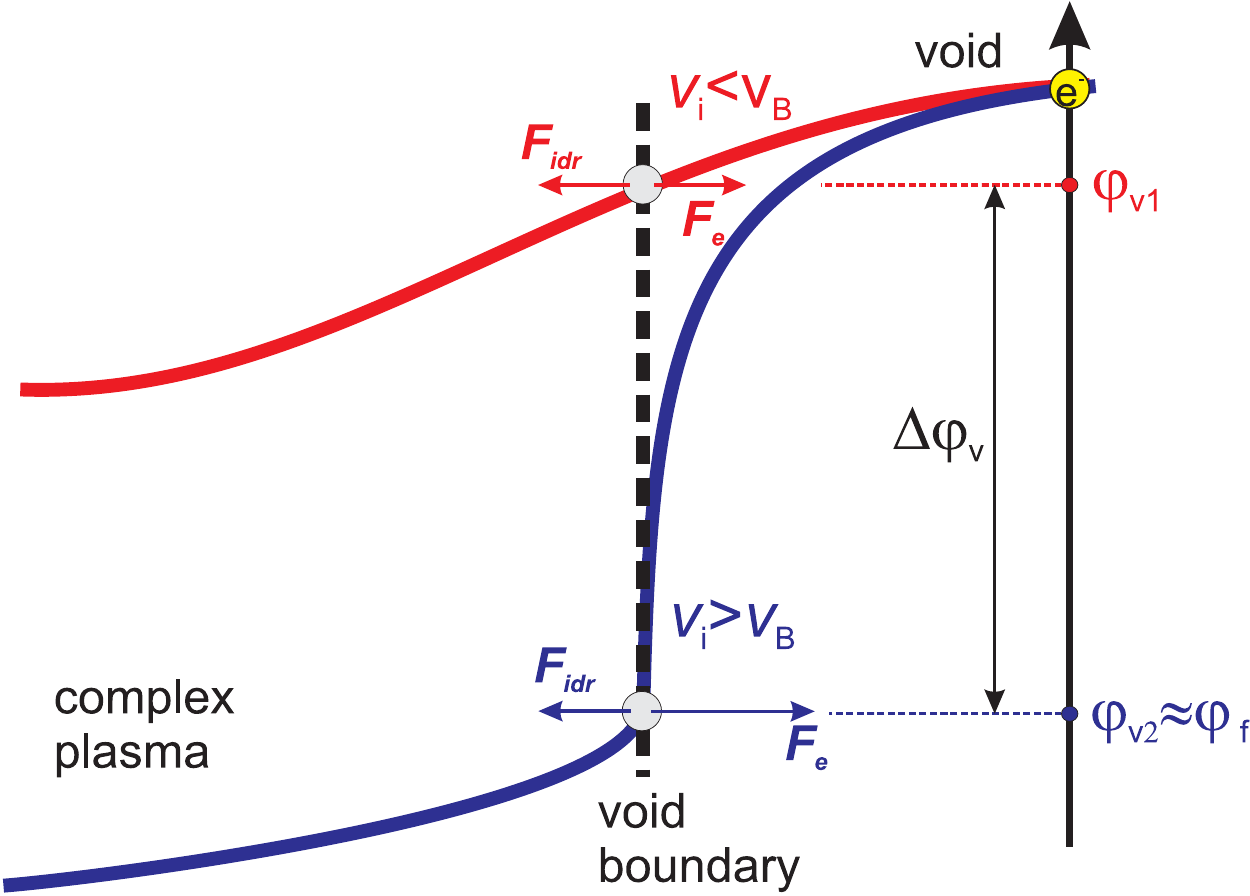}}
  \caption{Schematic representation of the sheath formation near the void boundary. Abrupt steepening of the electrostatic potential profile occurs when the ion velocity $v_i$ at the boundary exceeds the critical (Bohm) velocity $v_B$. This leads to the acceleration of the negatively charged microparticles towards the center of the void. The electrostatic potentials on the void boundary before and after the critical transformation are $\phi_{v1}$ and $\phi_{v2}$ respectively.}
  \label{Fi:SF}
\end{figure}

In low-pressure discharges the electron impact ionization is usually counterbalanced by the loss of electrons and ions on the electrodes and walls of the vacuum chamber. As pointed out above, microparticles immersed into a gas discharge provide an additional \textendash ~volumetric \textendash ~loss of charged particles. Loss rate on microparticles is proportional to their number density $n_d$.

Let us consider a boundary between a void and complex plasma with so small $n_d$, that plasma losses on the microparticles are negligible. In this case the electric field is simply determined by the ambipolar diffusion of charged particles to the electrodes and walls of the vacuum chamber. In an opposite limiting case $n_d$ can be considered so high that the presence of microparticles is equivalent to the placement of a solid floating electrode enclosing the void into the discharge. In this regime a sheath should be formed at the void boundary.

Transition from a smooth potential profile to formation of a sheath around the void should therefore occur at a finite value of $n_d$ when the ion loss rate on the microparticles is so large that the ions at the void boundary would have to be accelerated to a critical velocity (which, in an ideal case of a collisionless sheath, corresponds to the Bohm velocity $v_B$) in order to compensate it. Thus, our principal hypothesis is that the heartbeat instability occurs due to this critical phenomenon.

As illustrated in Fig.~\ref{Fi:SF}, sudden steepening of the electrostatic potential profile occurring at the critical transition has two important consequences. First, electric field at the void boundary increases dramatically, leading to the appearance of an uncompensated net force, which pushes the microparticles at the boundary towards the center. Second, in a transient situation of the sheath formation electron and ion fluxes through the void boundary are not necessarily equal. Electrons can therefore accumulate inside the void, causing the spatial redistribution of the glow intensity. Obviously, these consequences qualitatively explain the observed dynamics of microparticles and intensity of the plasma glow inside the void (Fig.~\ref{Fi:SE}b) during the heartbeat instability.

\subsection{Ionization balance inside the void}
Inside the void electrons and ions are produced by the electron impact ionization of neutral argon atoms, with the rate
\begin{equation}
Q_i=n\frac{p}{T}k_{gi}(T_e),
\label{Eq:Qi}
\end{equation}
where $k_{gi}$ is the rate constant of the electron impact ionization from ground state \cite{Gudmundson} and $n$ is the number density of bulk quasineutral plasma. The production rate of electrons and ions is compensated by their flux through the void boundary. Close to the critical condition of sheath formation the ionization rate should be equal to the Bohm rate of ion loss,
\begin{equation}
Q_B=n_{s}v_B\frac{S_{v}}{V_v},
\label{Eq:QB}
\end{equation}
where $n_{s}=ne^{-1/2}$ is the ion density at the sheath edge, $S_{v}$ is the surface area of the void boundary, $V_v$ is the volume of the void and $v_B=\sqrt{T_e/m_i}$ with $m_i$, being the argon ion mass, is the Bohm velocity. This yields
\begin{equation}
\frac{p}{T}k_{gi}(T_e)\sim\frac{1}{R_v}\sqrt{\frac{T_e}{m_i}},
\label{Eq:SF}
\end{equation}
where $R_v\equiv V_v/S_v$ is the effective void radius. Equation~(\ref{Eq:SF}) determines the electron temperature, at which the sheath on the void boundary should be formed. Substituting all values into Eq.~(\ref{Eq:SF}) we obtain $T_e=3.1$~eV for $p=20$~Pa.

The balance of ion and electron losses through the void boundary, $n_{s}v_B=1/4nv_{T_e}e^{-\phi_f/T_e}$, where $v_{T_e}=\sqrt{8T_e/\pi m_e}$ with $m_e$, being the electron mass, is the thermal velocity of electrons, determines its floating potential $\phi_f$, which is equal to $\approx 6.3T_e$ for argon \cite{Raizer}. In a steady state, void boundary acquires the potential $\phi_f$ since the net dc current is zero. However, evolution of the glow intensity in Fig.~\ref{Fi:SE} suggests that after the sheath is formed, transient currents should be induced in the system: void boundary should charge to a slightly larger negative potential (compared to $\phi_f$) in order to partly reject the electron flux from the void into the complex plasma.

Let us assume, that the electron temperature remains constant upon the critical transformation at the void boundary. The observed increase of the emission intensity inside the void can then only be attributed to the variation of electron density, which in turn can be expressed as follows:
\begin{equation}
\delta n_e=n\frac{p}{T}k_{gi}(T_e)\left[1-\exp{\left(-\frac{\delta\phi_f}{T_e}\right)}\right]\tau,
\label{Eq:dne}
\end{equation}
where $\delta\phi_f$ is the deviation of the potential of at the void boundary from $\phi_f$ and $\tau$ is the rise time of the emission intensity in Fig.~\ref{Fi:SE}b. Also,
\begin{equation}
\frac{\delta I}{I}=\frac{\delta n_e}{n},
\label{Eq:Deltas}
\end{equation}
where $\delta I$ is the observed increase of the emission intensity $I$. After combining Eqs.~(\ref{Eq:dne}) and~(\ref{Eq:Deltas}), we notice that electron density cancels out and $\delta\phi_f$ can be expressed in the following way:
\begin{equation}
\frac{\delta\phi_f}{T_e}=-\ln{\left[1-\frac{\delta I}{I}\frac{1}{\frac{p}{T}k_{gi}(T_e)\tau}\right]}.
\label{Eq:dphi}
\end{equation}
For the value of $T_e=3.1$~eV, obtained above, Eq.~(\ref{Eq:dphi}) yields $\delta\phi_f /T_e=8\times10^{-4}$. Therefore, even small deviations of the void boundary potential from the equilibrium (floating) value can cause the observed variations of the glow intensity.

Number density of $1s_5$ metastable state was found from the absorption measurements to be equal to $n_{1s_5}=1.6\times10^{16}$~m$^{-3}$. We can define the ratio of the rates of electron impact ionization out of the metastable and ground states,
\begin{equation}
s(T_e)=\frac{n_{1s_5}}{p/T}\frac{k_{mi}(T_e)}{k_{gi}(T_e)},
\end{equation}
where $k_{mi}$ is the rate constant of electron impact ionization from the metastable state, expression for which can be taken from Ref.~\onlinecite{Gudmundson}. For our case we obtain $s\sim10^{-4}$. Hence, ionization from metastable state is therefore consistently omitted in Eqs.~(\ref{Eq:Qi}),~(\ref{Eq:SF}) and~(\ref{Eq:dne}). It plays, however, the key role in the laser excitation of the heartbeat instability, making possible a small modulation of the ionization rate by the chopped laser beam (see Sec.~\ref{Subsc:LExHB}).

\subsection{Dynamics of microparticles near the void boundary}
According to our model, it is the same critical transformation of the electrostatic potential profile at the void boundary, which is responsible for both the increase of the plasma emission inside the void and the acceleration of microparticles towards the center of the void. Then the floating potential $\phi_f$ can be related to the dynamics of microparticles. After the formation of the sheath microparticles will be accelerated by the potential difference $\Delta\phi_v=\left|\phi_{v2}-\phi_{v1}\right|$, where $\phi_{v2}$ and $\phi_{v1}$ are the void potentials after and before the critical transformation, respectively (see Fig.~\ref{Fi:SF}), so that
\begin{equation}
Q_d\Delta\phi_v=\frac{m_dv^2}{2}.
\label{Eq:DPhi2}
\end{equation}
Here $m_d$, $v$ and $Q_d$ are the mass, velocity and charge of a microparticle, respectively. The microparticle charge can be expressed as follows \cite{RevIvlev}:
\begin{equation}
Q_d=4\pi\epsilon_0aT_ez,
\label{Eq:Q}
\end{equation}
where $z$ is a dimensionless factor depending on the ratio of electron and ion temperatures (and densities) as well as on the Mach number of streaming ions. We note that, according to the Bohm criterion \cite{Riemann}, the sheath forms when the drifting ions reach the energy of $\phi_B=T_e/2$. Therefore, $\phi_{v1}$ cannot significantly exceed the value of $\phi_B$. Then $\phi_{v1}$ (and, of course, $\delta\phi_f$) is substantially smaller than $\phi_f \approx 6.3T_e$ and, consequently, $\Delta\phi_v$ is somewhat smaller than $\phi_f$. Combining Eqs.~(\ref{Eq:DPhi2}) and~(\ref{Eq:Q}) we obtain
\begin{equation}
v\simeq\sqrt{\frac{8\pi\epsilon_0 a T_e z \phi_f}{m_d}}.
\label{Eq:v}
\end{equation}
Then, after the substitution of the corresponding parameters in Eq.~(\ref{Eq:v}) and assuming $z=1$ deep inside the sheath we obtain $v\approx1.5$~m/s.

We measured the velocity of the accelerated microparticles averaged over about $5$~ms after they start their motion into the center of the void. Effect of the Epstein neutral drag \cite{Epstein}, whose characteristic time for the conditions of our experiment is $13$~ms, is then negligible. According to the measurements $v\approx 0.5$~m/s, which is not inconsistent with the value obtained from Eq.~(\ref{Eq:v}).

\subsection{Dependence on the discharge parameters}
We supposed that the heartbeat instability occurs due to the formation of the sheath at the void boundary. This happens when the diffusion rate of the plasma through the void boundary, $Q_{diff}\sim n D_a/R_v^2$ (where $D_a\propto T_e/p$ is the ambipolar diffusion coefficient), reaches the value of $Q_B$ determined by Eq.~(\ref{Eq:QB}). Due to the ionization balance $Q_{diff}$, should always be equal to $Q_i$, determined by Eq.~(\ref{Eq:Qi}), which yields the following scaling:
\begin{equation}
pR_v\propto\sqrt{T_e/k_{gi}(T_e)}.
\label{Eq:rpSca}
\end{equation}
We note that this should not be understood as a self-consistent scaling for the void size, since the balance of forces acting on the microparticles on the void boundary is not taken into account in Eq.~(\ref{Eq:rpSca}).  Next, we consider the ratio
\begin{equation}
C=\frac{Q_{diff}}{Q_B}=\frac{Q_i}{Q_B}\propto\frac{\sqrt{T_e}}{pR_v}\propto{\sqrt{k_{gi}(T_e)}},
\label{Eq:Sca}
\end{equation}
which controls the formation of the sheath at the void boundary and, hence, the appearance of the heartbeat instability. The rhs. of Eq.~(\ref{Eq:rpSca}) is a decreasing function of $T_e$ (at least while $T_e$ is less than the ionization potential). This means, that the ratio $C$  decreases with the product $pR_v$.

Scaling~(\ref{Eq:Sca}) can be used to understand \textit{e.g.} certain features of the stability diagram shown in Fig.~\ref{Fi:HBPD}. According to Ref.~\onlinecite{Tatanova} $T_e$ monotonically decreases with the discharge power in a microparticle-free rf discharge. This can explain the existence of the upper power threshold for heartbeat instability. On the other hand, monotonic decrease of $T_e$ with the discharge power contradicts the existence of the lower power threshold. However, in a discharge with microparticles this dependence could easily become non-monotonic due to \textit{e.g.} variable microparticle charge or size of the void. Furthermore, since the parameter $C$ decreases with $pR_v$, presence of a pressure threshold in Fig.~\ref{Fi:HBPD} becomes also clear. Increase of $p$ suppresses the ambipolar diffusion, so that above some critical value of pressure the sheath on the void boundary can be no longer sustained irrespective of the rf power.

In the framework of this approach, the onset of the instability upon the increasing of the number of levitating particles can also be explained: Injecting additional microparticles into the discharge leads to the decrease of the void size $R_v$ and, therefore, increase of the ratio $C$.

\subsection{Effect of the tunable laser beam}\label{Subsc:LExHB}
The laser-excited instability is not the result of the radiation pressure exerted on the microparticles. First of all, the instability was observed at the laser intensity of $J$ = $600$~W/m$^2$, which is hardly enough to produce significant dynamical effects \cite{RadPress}. Also, the instability disappears once the laser wavelength is detuned from the center of the absorption line. The effect produced by the laser is, therefore, due to its interaction with the background plasma.

\begin{figure}[t!]
\centering
  \centerline{\includegraphics [width=3.in]{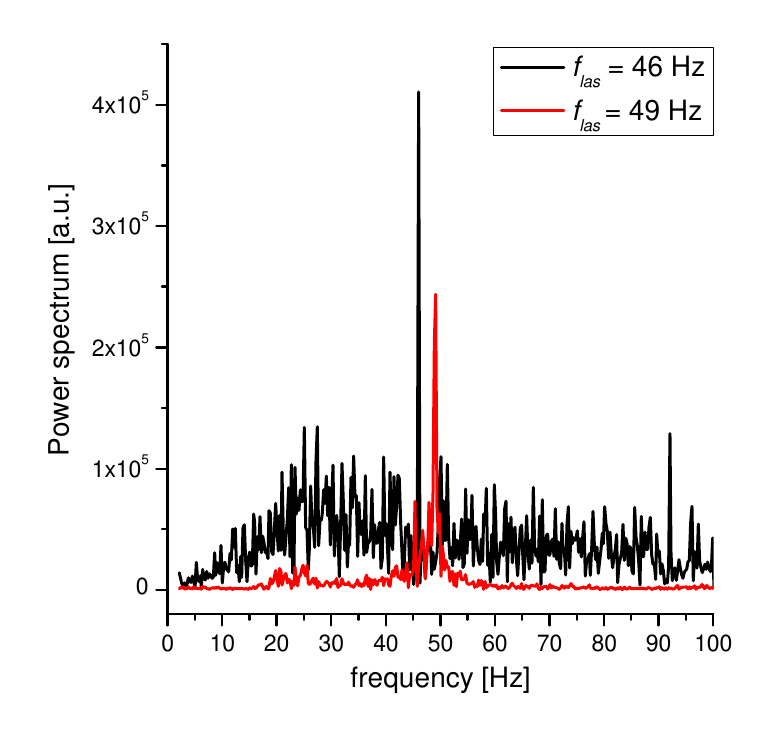}}
  \caption{Fourier spectra of the light intensity inside the void, for the experimental points from Fig.~\ref{Fi:LEHBDep}a. At $f_{las}=46$~Hz (when the intensity of the heartbeat is close to maximum) the spectrum is peaked at about the half of the laser frequency, suggesting the parametric instability as underlying mechanism of the laser-excited heartbeat.}
  \label{Fi:Spectra}
\end{figure}

The fact that the instability can only be triggered when the laser beam hits the void indicates the minor importance of any kind of direct interaction between metastable atoms and microparticles (e.g., secondary electron emission \cite{MetEmission}). Therefore, in case of laser excitation it is only the perturbation of the ionization that can lead to the instability. Ionization from metastable states provides only a small fraction $s\sim 10^{-4}$ to the net ionization rate. This suggests that the observed resonance is the consequence of the parametric instability. To address this issue we analyzed the Fourier spectra of signals from the photomultiplyer, pointed to the void (Fig.~\ref{Fi:Spectra}).
Close to the resonance at $f_{las}=46$~Hz, when intensity of the heartbeat is high, we observed excitation of a broad frequency band with the peak in the vicinity of $23$~Hz, \textit{i.e.}, halved driving frequency. This is one of the characteristic features of a parametric instability. At $f_{las}=49$~Hz, when no heartbeat is observed, this band is very much suppressed, as compared to the case of $f_{las}=46$~Hz. Due to the parametric instability the plasma parameters inside the void could be changed in such a way, that the sheath formation at the void boundary could be triggered.

Next, we show that the laser beam with the intensity $J$ tuned to the wavelength of $2p_7\to 1s_5$ transition of argon can significantly disturb the balance of metastable atoms. The laser light is absorbed by the $1s_5$ argon metastable atoms is absorbed with the rate
\begin{equation}
Q_{abs}=\Gamma\sigma_{abs} n_{1s_5}.
\label{Eq:AbsRate}
\end{equation}
Here $\Gamma=\lambda J/{hc}$ is the flux of laser photons and
\begin{equation}
\sigma_{abs}=\frac{\lambda}{8\pi}\sqrt{\frac{m}{2T}}\frac{g_{2p_7}}{g_{1s_5}}A.
\label{Eq:PhotAbs}
\end{equation}
is the absorption cross-section, where, $g_{2p_7}=3$ and $g_{1s_5}=5$ are the statistical weights of the respective atomic states and $A=5.18\times 10^6$ s$^{-1}$ is the Einstein coefficient of the transition \cite{Wiese}. Equation~\ref{Eq:PhotAbs} holds only for the center of a spectral line, with a Doppler profile characterized by the temperature $T$. Absorption measurements, performed with the same laser and using the same transition, showed that the spectral line indeed has a Doppler shape with $T\approx$ $300$~K. By substituting these numbers into Eqs.~(\ref{Eq:AbsRate}) and~(\ref{Eq:PhotAbs}), we get $Q_{abs}$ = $5.6\times10^{21}$~m$^{-3}$s$^{-1}$. Branching fractions for the atomic level 2p$_7$ of argon are such that only $23$\% of atoms, excited to this level return into metastable states \cite{Wiese}. The rest will decay to 1s$_2$ and 1s$_4$ radiative states, whose lifetimes are orders of magnitude shorter than those of the metastable states. Therefore, our laser beam produces an effective sink of metastable atoms.

For comparison we estimate the rate of electron impact excitation from ground state into 1s$_5$ metastable state for the typical value of electron density $n_e$ = $10^{14}$ m$^{-3}$ and temperature $T_e$ = $3.1$~eV (obtained from the ionization balance). The rate of this process, $Q_{exc}=3.4\times10^{19}$ m$^{-3}$s$^{-1}$ \cite{Gudmundson} is much smaller than the rate of destruction of metastable states by the laser beam. Number density of metastable states inside the laser beam is significantly reduced compared to the unperturbed plasma.

\section{Conclusion}
The heartbeat instability is identified as a heterogeneous phenomenon, occurring when a clear interface exists between the microparticle-free region (void) and complex plasma. In the proposed simple model, the instability occurs due to the formation of the sheath on the void boundary. Abrupt change of the electrostatic potential profile leads to the appearance of uncompensated electrostatic force pushing the microparticles towards the center of the void. Transient variations of the potential of the void boundary can be responsible for the observed variations of the plasma glow intensity in the entire discharge. We have also shown that our hypothesis can be useful in understanding the effect of different parameters (\textit{i.e.} discharge power, gas pressure and number of levitating microparticles) on the heartbeat instability. Excitation of the heartbeat by a modulated tunable laser beam was found to be the result of a parametric instability.

To conclude, we would like to stress that our hypothesis on the mechanism of the heartbeat instability should not be considered as a self-consistent approach, aiming to completely solve the problem. Clearly, careful dedicated experiments are needed to further confirm or deny this hypothesis.

\section{Acknowledgements}
We acknowledge the funding by the space agency of the Deutsches Zentrum f\"{u}r Luft- und Raumfahrt e.V. with funds from the federal ministry for economy and technology according to a resolution of the Deutscher Bundestag under grant No. (50WP0203)

\end{document}